\documentclass[10pt,conference]{IEEEtran}

\usepackage{cite}
\usepackage{epsfig}
\usepackage{times}
\usepackage{graphicx}
\usepackage[figuresright]{rotating}
\usepackage{subfigure}
\usepackage{amssymb}
\usepackage{amsmath}
\usepackage{setspace}
\usepackage{rotating}
\usepackage{multirow}
\usepackage{wrapfig}
\usepackage{booktabs}
\usepackage{array}
\usepackage{comment}
\usepackage{subfigure}
\usepackage{epstopdf}
\IEEEoverridecommandlockouts

\usepackage{algorithmic}
\usepackage{algorithm}
\floatname{algorithm}{Algorithm} \textfloatsep 0.5cm

\usepackage{colortbl}

\begin{document}
\title{\huge{Towards Low-power Wearable Wireless Sensors for Molecular Biomarker and Physiological Signal Monitoring}}
\author{Xueyuan Zhao, Vidyasagar Sadhu, Tuan Le, Dario Pompili, Mehdi Javanmard\\
Department of Electrical and Computer Engineering, Rutgers University--New Brunswick, NJ, USA\\
Emails: \{xueyuan.zhao, vidyasagar.sadhu, tuananh.le, pompili, mehdi.javanmard\}@rutgers.edu}

\maketitle

\linespread{0.98}

\thispagestyle{empty}
\pagestyle{plain} \pagenumbering{arabic}
\pagestyle{empty}

\begin{abstract}
A low-power wearable wireless sensor measuring both molecular biomarkers and physiological signals is proposed, where the former are measured by a microfluidic biosensing system while the latter are measured electrically. The low-power consumption of the sensor is achieved by an all-analog circuit implementing Analog Joint Source-Channel Coding~(AJSCC) compression. The sensor is applicable to a wide range of biomedical applications that require real-time concurrent molecular biomarker and physiological signal monitoring.
\end{abstract}

\section{Introduction}\label{sec:introduction}
Current technologies for monitoring individuals' physiological signals including Galvanic Skin Response~(GSR), Electrocardiogram~(ECG), and Electroencephalogram~(EEG) have the major limitation that the signals only reflect the tissue-level electronic phenotype of human beings. The measurement of small molecules and macromolecules such as nucleic acids, proteins, and cells can provide a more accurate means for understanding the biomolecular pathways inside the human body, which is invaluable for understanding patient health. The use of integrated and embedded computing power has enabled wearable electronic systems for monitoring various physiological signals~\cite{Konijnenburg16}. Meanwhile, we have also demonstrated the detection of various biomolecules in blood using impedance-based biodetection, which has the potential to enable truly miniaturized wearable diagnostic platforms~\cite{Lin15}. The combination of wearable devices for continuous biomolecular measurements and physiological parameter acquisition can provide a more accurate understanding of the physiological state of millions of individuals and enable the new paradigm of ``crowd-sourced" biomarker discovery and validation.

\textbf{Related Work:}
Various bodily fluids such as blood and sweat can provide a plethora of information regarding the physiological state of an individual. Existing wearable wireless sensors including hybrid biosensors are all based on digital circuits for signal processing and wireless transmission. Various groups have developed wearable systems for monitoring biomarkers in sweat. A sensing system with sweat-lactate concentration measured by immobilized enzyme with concurrent ECG measurement was previously demonstrated in~\cite{Imani16}, where the sweat-lactate concentration and ECG signal are first passed to Analog-to-Digital Converters~(ADCs) for the processing by a microcontroller, and then are wirelessly transmitted by a Bluetooth module. An integrated sensor is designed in~\cite{Gao16} with hybrid biomarkers detection and skin temperature measurement. The measurement signals are routed to a conditioning circuit, sampled with an ADC, processed by a microprocessor, and then sent to a mobile phone wirelessly via a Bluetooth interface. In~\cite{Luhmann16}, a wireless sensor that measures the EEG and neurophysiological signal is proposed for digital Bluetooth transmission. In~\cite{Abrar16}, the sweat lactate is measured and the sensing signal is sampled with an ADC, processed by a microprocessor, and then transmitted by a digital Near-Field Communication~(NFC) wireless module. In~\cite{DeHennis16}, a wireless sensor to measure glucose was demonstrated and the digital NFC communication was adopted for wireless transmission. In all these systems, the major power consumption occurs is due to the digital conversion by the ADCs.

\begin{figure}
\begin{center}
\includegraphics[width=3.2in]{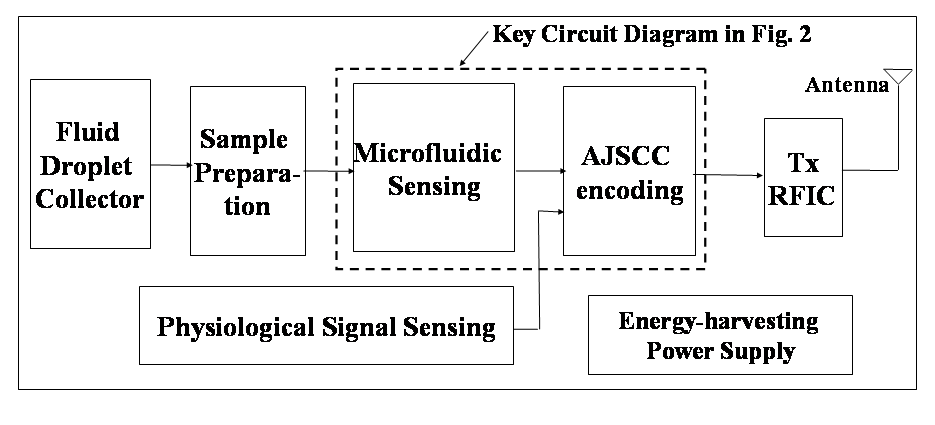}
\end{center}
\vspace{-0.15in}
\caption{All-analog wireless wearable sensor for real-time dual measurement.}\label{fig_proposed_sensor}
\vspace{-0.15in}
\end{figure}

\begin{figure*}
\begin{center}
\includegraphics[width=6in]{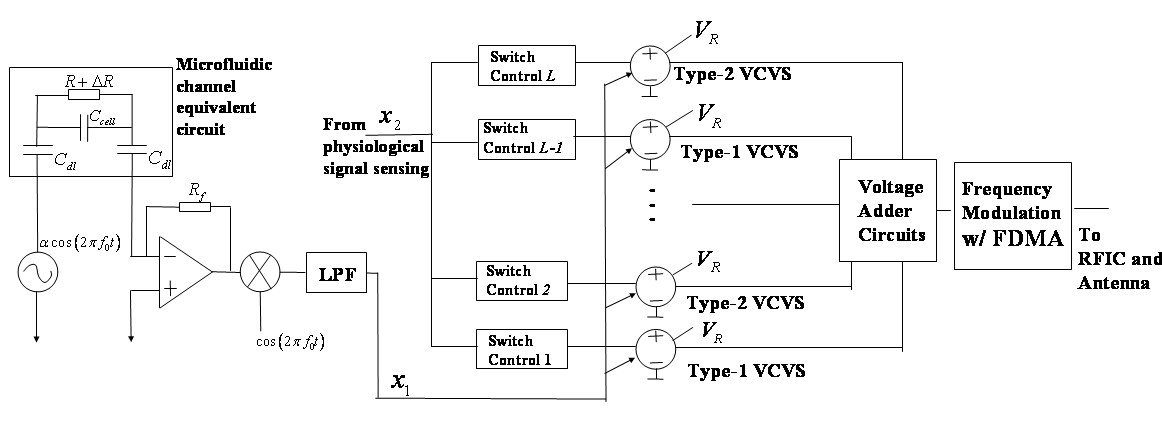}
\end{center}
\vspace{-0.15in}
\caption{All-analog circuit diagram with microfluidic and physiological sensing signal compressed by Analog Joint Source Channel Coding~(AJSCC) circuit~\cite{Zhao16} with Frequency Division Multiple Access~(FDMA) in order to multiplex a large number of sensors.}\label{fig_Proposed_Biosenor}
\vspace{-0.15in}
\end{figure*}

\textbf{Our Approach:}
We propose a new type of all-analog wireless circuit that can replace the digital wireless transmission of the wearable wireless sensors. In our approach, both biomarker measurements from a microfluidic biosensor and physiological signals are compressed into a single signal using Analog Joint Source Channel Coding~(AJSCC)~\cite{Zhao16} and transmitted wirelessly to a digital Cluster Head~(CH) receiver that does the AJSCC decoding to recover the source signals (Fig.~\ref{fig_proposed_sensor}). All of the signal processing at the transmitter happens \emph{in the analog domain} without the need for power-hungry ADC/DAC components, therefore saving power (more so in case of high-frequency sensing scenarios). The proposed sensor has a wide range of biomedical applications. For example, monitoring of neuro-degenerative diseases such as Alzheimer's disease can benefit from measuring jointly the physiological signal of EEG and neuro-related biomarkers. Another example pertains to the continuous monitor of cardiovascular disease, where the physiological signal of ECG and blood biomarkers such as cardiac troponin or creatanine should be acquired. The power consumption can be kept low at the sensor node due to the fact that there are no ADC/DAC components in the system. All the analog components in AJSCC can be integrated into an analog Integrated Circuit~(IC), thus the power consumption can be kept much lower than using digital IC with ADC/DACs. The power consumption of the proposal is compared under the same assumption against an IC design and shown to have a consumption of $130~\mathrm{\mu W}$~\cite{Zhao16}, which is much lower than that of a typical microprocessor chipset requiring a few $\mathrm{mW}$.

\textbf{Outline:}
In Sect.~\ref{sec:sensor}, our hybrid biosensor for dual measurements is proposed; in Sect.~\ref{sec:exp}, the circuit experimental results are presented; finally, in Sect.~\ref{sec:conc}, conclusions are drawn.

\section{Proposed Solution}\label{sec:sensor}
We are proposing an all-analog wearable wireless sensor for real-time dual measurement---the  molecular biomarker concentration and physiological signals. The proposed sensor (Fig.~\ref{fig_proposed_sensor}) is composed of the following blocks: a fluid droplet collector, sample preparation, microfluidic sensing, physiological signals sensing, AJSCC compression, RFIC, antenna, and energy-harvesting power supply. The molecule biomarker can be measured from bodily fluids such as blood, sweat, or saliva. 

\textbf{Microfluidic Sensing:}
The fluid droplet collector pumps up the fluid for measurement; this way the fluid is sent to the sample preparation to be purified so as to obtain a fluid with target molecule to measure. The fluid flow is further sent to the microfluidic sensing with the analog readout circuit, as depicted in Fig.~\ref{fig_Proposed_Biosenor}. The molecules passing the electrode will produce pulses in the voltage output as described in~\cite{Lin15,Emaminejad12}. The physiological signal sensing continuously acquires signals from electrical measurement. The output signals of physiological and microfluidic sensing are sent to the AJSCC encoding, which compresses efficiently two analog signals into one, as detailed in our previous work~\cite{Zhao16}. The compressed signal is then passed to the RFIC and antenna for radio transmission. The whole system is very energy efficient and can therefore be fully powered by energy-harvesting power supply (i.e., it may utilize human motion to generate electricity as in~\cite{Wang2016}), with no need for duty cycles/sleep modes. The whole sensor can be fabricated into an integrated system to ensure the ultra-compact platform footprint necessary for wearable applications. The key features of our proposal are: i) low-power consumption thanks to an all-analog circuit realization where no power-hungry ADCs are used and ii) hybrid continuous monitoring. 

\textbf{All-analog Wireless Sensor Circuit:}
The proposed key circuit is depicted in Fig.~\ref{fig_Proposed_Biosenor}. The fluid sample are pumped into the system for sample preparation, then sent to the microfluidic channel. The microfluidic system adopts impedance-based molecule-concentration measurement. The electrodes are fabricated inside the microfluidic channel for impedance detection. The equivalent circuit inside the microfluidic channel can be easily modeled by two double-layer capacitors $C_{dl}$, and one resistor $R$ and one capacitor $C_{cell}$ in parallel. When a molecule passes by, the resistor will have a change in resistance $\Delta R$, which will result in one pulse in the output impedance detection signal. The impedance measurement is excited with a cosine wave with frequency $f_0$. The resistance change $\Delta R$ and the voltage change will be detected and amplified by the lock-in operational amplifier with feedback resistance $R_f$. The signal is then passed to a mixer with frequency $f_0$ and then passed through a low-pass filter. Finally, the impedance signal is input to the AJSCC encoding circuit.

\textbf{Transmitter Signal Compression:}
In the all-analog realization of AJSCC circuit~\cite{Zhao16}, the microfluidic sensing signal $x_1$ is the signal to control the output of Voltage Controlled Voltage Sources~(VCVS). Each VCVS is switched between saturation voltage $V_R$, the linear voltage output corresponding to $x_1$, and ground. The physiological signal $x_2$ is the control signal of the VCVS composed of $L$ stages. With higher physiological signal, there will be more stages being activated. For example, if there is $M$ stages being activated, the $M$-th stage will be controlled by microfluidic sensing signal $x_1$ to produce a continuous varying output voltage, and the $1$-st to the $M-1$-th stages will output the maximum voltage $V_R$. The other higher $L-M$ stages are producing zero grounded outputs. The voltages of all the stages are summed together by an analog voltage adder to produce the desired AJSCC encoded voltage. The signal is then passed to RFIC and the antenna. 

The advantage of this circuit is that the sensor power consumption can be kept much lower than using a digital board. For a digital system, the power consumption is high due to the digital microcontroller and other digital devices in the system, including the Analog-to-Digital Converters~(ADC) and microprocessor. The power consumption of our analog AJSCC board without radio power is estimated to be $130~\rm{\mu W}$ for a discrete-component realization~\cite{Zhao16}, and can be even lower (less than $50~\rm{\mu W}$) if an IC design is adopted. For state-of-the-art wireless digital sensors (e.g.,~\cite{wsn430,telosb}) the consumption is at least $1$-$2~\rm{mW}$ in active mode. A detailed power comparison of our AJSCC system with existing wireless sensors can be found in~\cite{wons3tier2017}. The fabrication cost can also be kept lower than the digital counterpart leveraging mass IC production.

\textbf{Receiver Signal Processing:}
The received RF signal from antenna is firstly downconverted to baseband by the receiver RFIC, and then sampled by ADC with sampling frequency $f_s$ using a National Instruments~(NI) Digital Acquisition~(DAQ) device. The digital output from the ADC of NI DAQ device are sent to NI~LabView on a host computer. Fast Fourier Transform~(FFT) of size $N_s$ is performed on the samples; then, the frequency peaks are detected to recover the frequency positions for frequency-modulated signal and the two compressed signals are obtained by mapping back to the AJSCC point in LabView on the host computer. AJSCC decoding is performed at the receiver to recover the original mapped point. Further Signal-to-Noise Ratio~(SNR) improvement by filtering can be performed on the recovered signal.

\begin{figure}
\begin{center}
\includegraphics[width=3.5in]{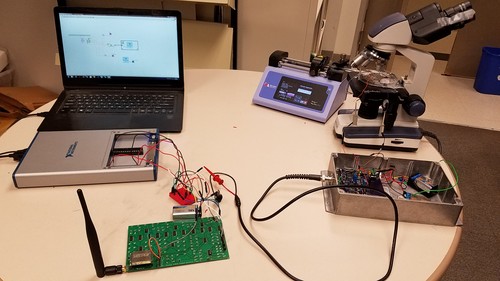}
\end{center}
\vspace{-0.15in}
\caption{The right-hand side shows the cytometry system, which generates the impedance cytometry signal. Prior collected GSR data is used to generate an analog signal using NI~LabView/DAQ system~(left top). Both these analog signals are fed to the AJSCC sensor~(left bottom), which does the AJSCC encoding and transmits wirelessly to a digital CH receiver (not shown).}\label{fig_system}
\vspace{-0.15in}
\end{figure}

\section{Experimental Results}\label{sec:exp}
The experiment is designed as a proof of concept to prove the functionalities of the key circuit at the transmitter side as well as of the receiver to recover the two compressed signals. 

\textbf{Experimental Setup:}
The test is performed in an indoor environment with a carrier frequency of $2.4~\rm{GHz}$ and an omni-directional antenna. The distance between the transmitter and receiver is about 5 meters. The transmitter system is shown in Fig.~\ref{fig_system}. The microfluidic signal (which is chosen to be a micro-impedance cytometry signal for proof-of-concept demonstration) is generated by micron-sized beads passing through a channel with a width of $30$~microns and height of $20$~microns. The signal is excited with cosine-ware frequency $f_0$ of $500~\rm{kHz}$. The flow rate is $0.1$~micro Liter/min. The bead site is $7.8$~microns to mimic a blood cell. A sequence of pulses is generated with beads passing electrode. The physiological signal is chosen to be Galvanic Skin Response~(GSR) signal for the experiment purpose, which measures the skin conductance and is instrumental in assessing physiological stress. Since GSR data collection is done via Bluetooth using Shimmer sensors, the collected data is used to generate an analog signal using NI LabView/DAQ system. Both these signals are fed to the circuit board developed in~\cite{Zhao16} to perform the sensor signal encoding in the analog domain. In that system, we have given microfluidic signal to the x-dimension (which is continuous) and physiological signal to the y-dimension (quantized to 10 levels).

\begin{figure}
\begin{center}
\includegraphics[width=3.5in]{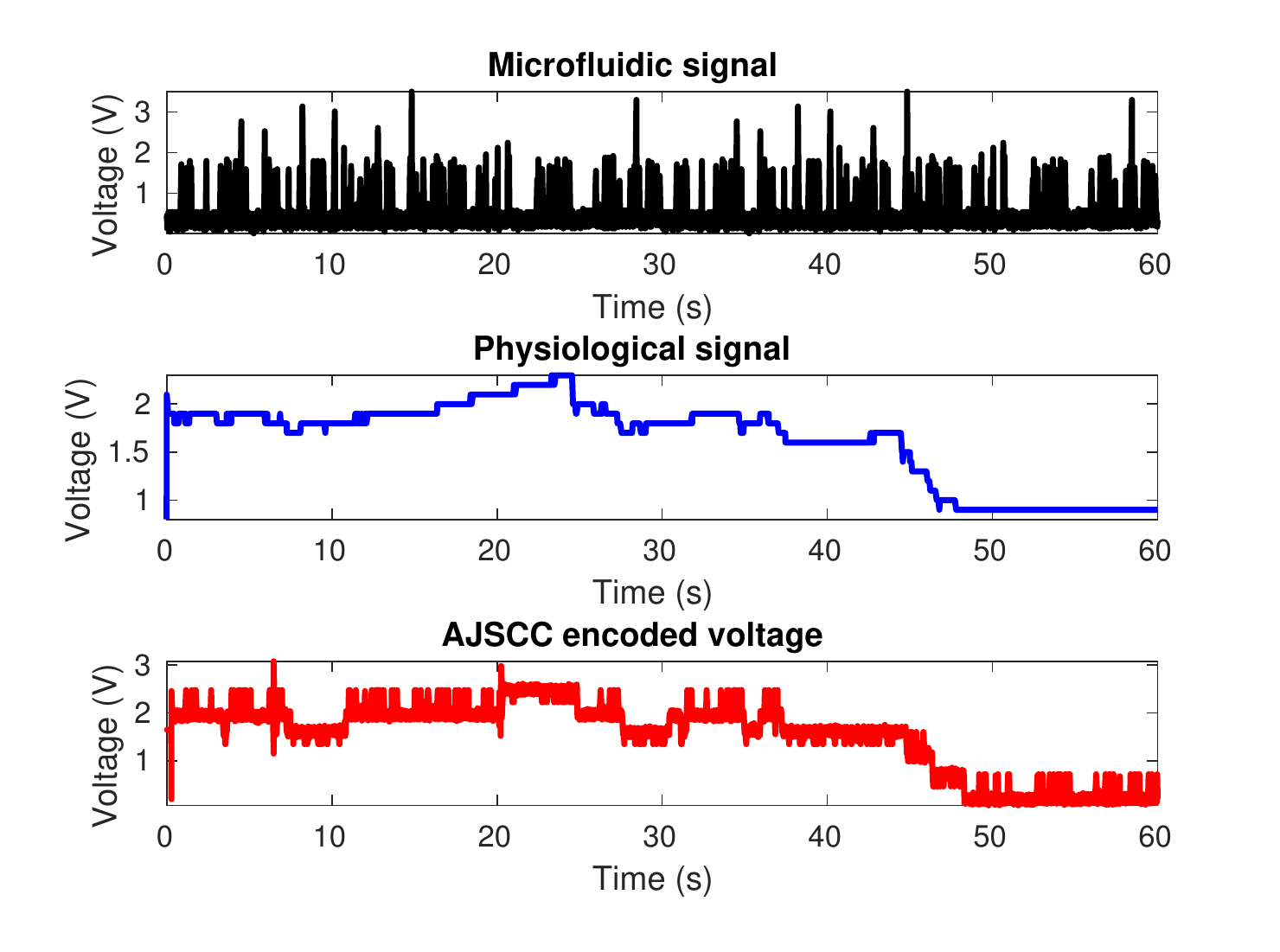}
\end{center}
\vspace{-0.15in}
\caption{The three signals: microfluidic (cytometry), physiological (GSR), and AJSCC-encoded voltage, at the transmitter.}\label{fig_output_Tx}
\vspace{-0.15in}
\end{figure}

\textbf{Transmitter AJSCC Encoding:}
The original microfluidic signal, physiological signal, and the AJSCC encoded signal in the transmitter are depicted in Fig.~\ref{fig_output_Tx} as captured using NI DAQ hardware and LabView software. 
We can observe that AJSCC encoding compresses the two signals into a single one such that the microfluidic signal is riding on top of the physiological signal. We note that our AJSCC sensor board introduces a quantization error in the physiological signal since it has only 11 levels in the y-dimension, as seen in Fig.~\ref{fig_output_Tx} where the physiological signal portion in the AJSCC encoded signal has been quantized. This AJSCC encoded signal is frequency modulated, upconverted, and then transmitted using Commercial Off The Shelf~(COTS) RFIC chip. %

\begin{figure}
\begin{center}
\includegraphics[width=3.5in]{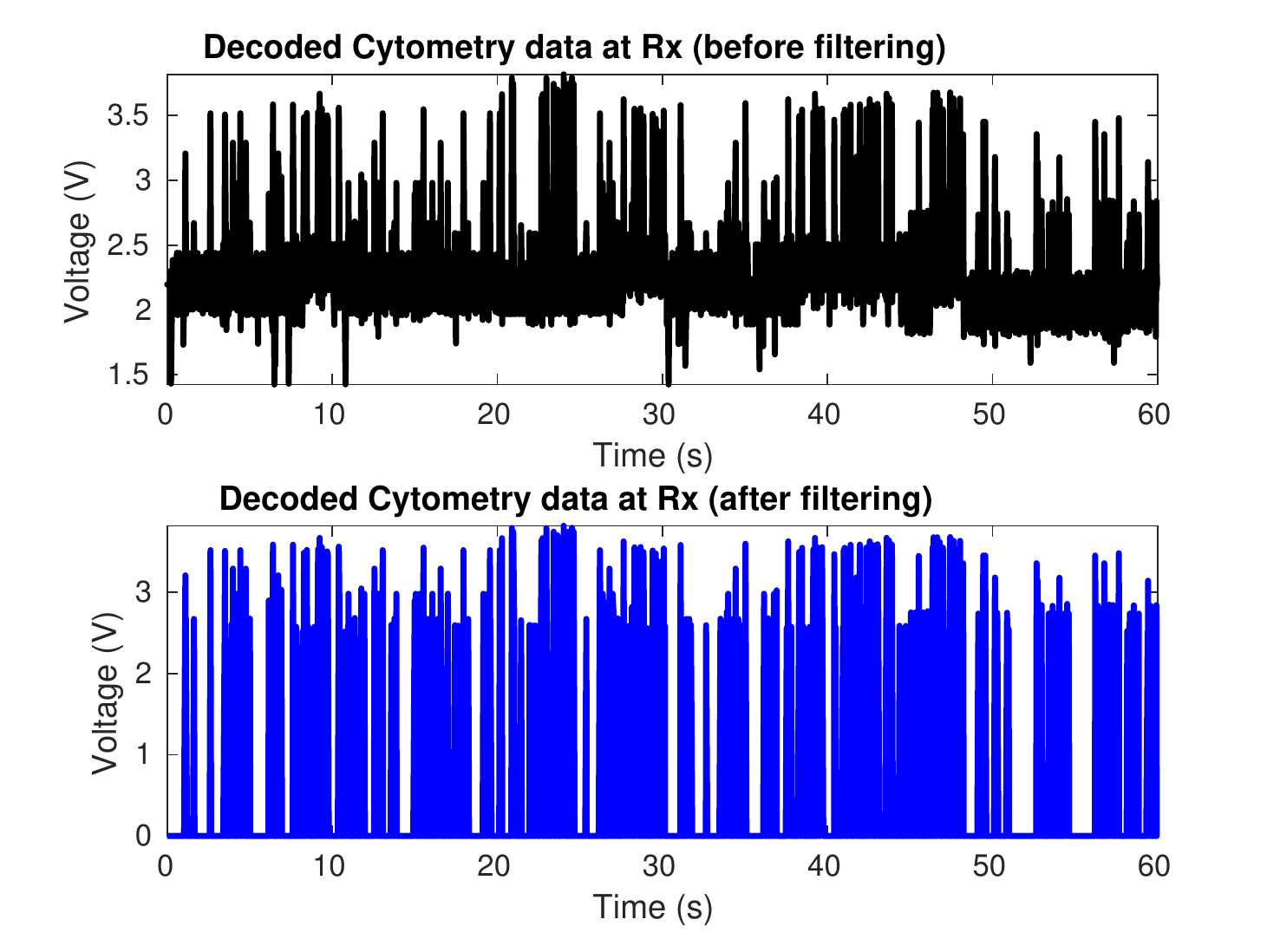}
\end{center}
\vspace{-0.15in}
\caption{Decoded microfluidic (cytometry) signal in the CH receiver---before (top) and after (bottom) filtering.}\label{dec_cytometry}
\vspace{-0.15in}
\end{figure}

\begin{figure}
\begin{center}
\includegraphics[width=3.5in]{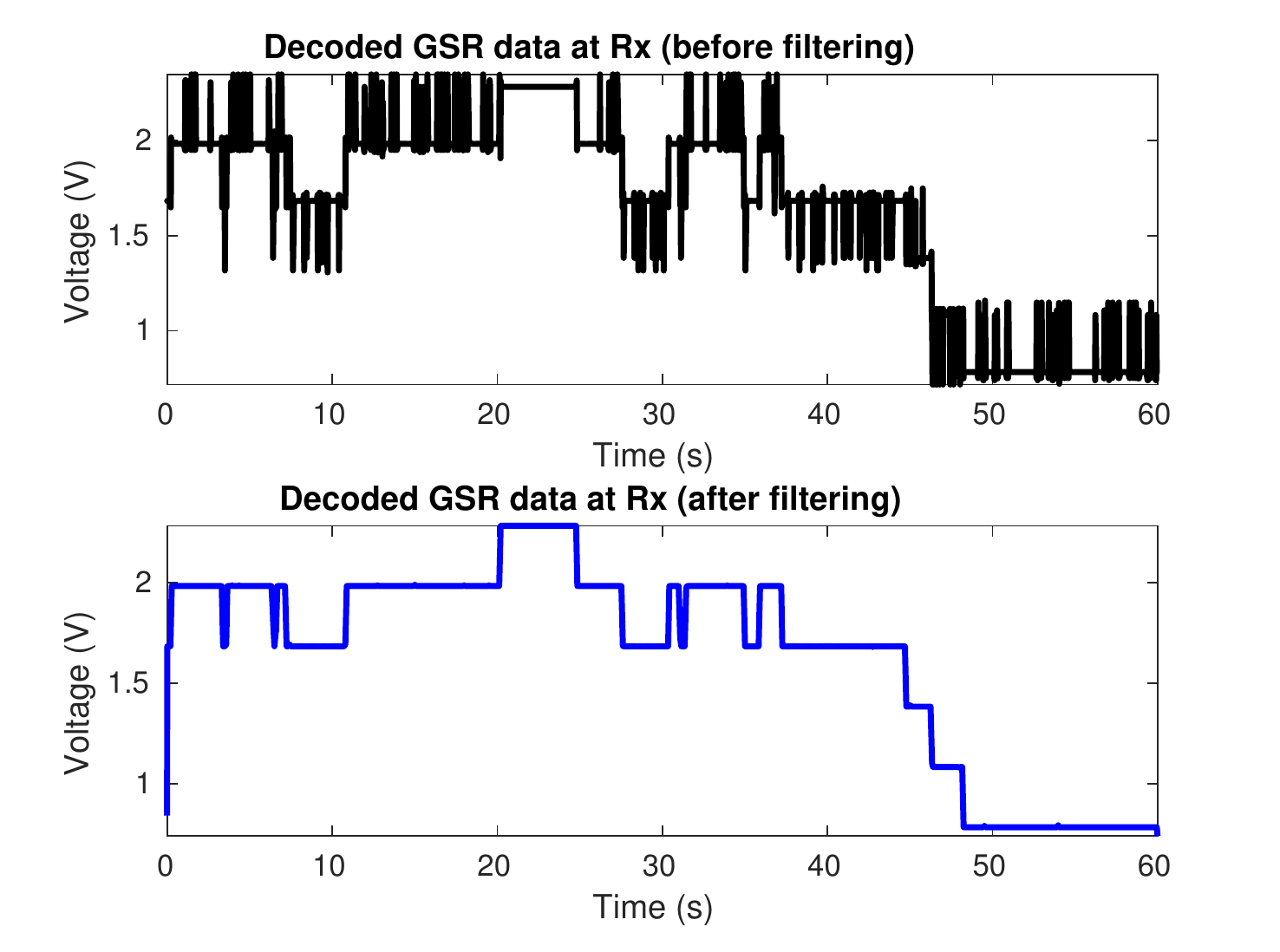}
\end{center}
\vspace{-0.15in}
\caption{Decoded physiological~(GSR) signal in the CH receiver---before (top) and after (bottom) filtering.}\label{fig_dec_gsr}
\vspace{-0.15in}
\end{figure}

\textbf{Receiver Decoding:} 
In the receiver, the parameter $N_s$ is chosen as $5,000$ and sampling rate $f_s$ is set to $500~\rm{kHz}$, which is the maximum supported by the DAQ device. AJSCC encoded signal is recovered from the frequency values using simple linear mapping (as done in the transmitter board). The AJSCC encoded voltage is then decoded to individual physiological and microfluidic signals using simple modulo arithmetic~\cite{Zhao16}. The decoded microfluidic signal in LabView is shown in Fig.~\ref{dec_cytometry}~(top). It can be observed that the peaks of the signal are recovered. The error at the bottom of the signal is due to the bias in the AJSCC stage mapping. The decoded physiological signal is shown in Fig.~\ref{fig_dec_gsr}~(top). The quantization effect of the recovered physiological signal is due to the fact that in the transmitter board only 11 stages are designed for the AJSCC mapping. The error floor at the bottom of the microfluidic signal is removed using a thresholding filter, where the threshold value is set just above the error floor level. The unwanted spikes in the physiological signal are smoothed using a $200$-th order median filter. The bottom portions of Figs.~\ref{dec_cytometry} and \ref{fig_dec_gsr} show the results after filtering.

In the experiments, the NI~DAQ device was sampling the received baseband signal at $500~\rm{KHz}$; however, the NI LabView was not able to process the data at the same rate. Hence, we collected the samples alone in the LabView (without further processing) and then processed the samples in MATLAB for frequency detection and AJSCC decoding. This also gave us the opportunity to vary the window size $N_s$ (and find the best value) for frequency detection: a small window means fewer samples for frequency estimation resulting in wrong frequency values; on the other hand, a large window results in loss of sampling accuracy (as we get averaged values instead of actual values). %
After trying different values, we found that $N_s$=$5,000$ gives the best results in that it matches closely the original physiological and microfluidic signals.

\section{Conclusions and Future Work}\label{sec:conc}
We proposed a low-power all-analog wearable wireless sensor, applicable to a wide range of biomedical applications, implementing Analog Joint Source-Channel Coding~(AJSCC) compression for continuous molecular biomarkers and physiological signal monitoring, where the former are measured by a microfluidic biosensing system while the latter are measured electrically.
We are now combining micro-needles with the microfluidic sensor to devise a minimally-invasive method for extracting blood from an individual and measuring multiple types of molecular biomarkers and physiological parameters. %

\textbf{Acknowledgments:}
We thank ECE students A.~Padwad, K.~Ahuja, and N.~Maiya for their help with the experiments.

\bibliographystyle{IEEEtran}\small
\bibliography{ref_cytometry_ajscc}

\end{document}